\documentclass[twocolumn,showpacs,superscriptaddress,amsmath,amssymb]{revtex4}
\usepackage[utf8]{inputenc}
\usepackage{epsf}
\usepackage{graphicx}
\usepackage{latexsym}
\usepackage{amsmath}
\usepackage{amssymb}
\usepackage{amsfonts}
\usepackage{color}
\begin{document}


\title{Nonuniform quantum--confined states and \\ visualization of hidden defects in thin Pb(111) films}
\author{A.\,V. Putilov$^a$, S.\,S. Ustavschikov$^{a,b}$, S.\,I. Bozhko$^c$ and A.\,Yu. Aladyshkin$^{a,b}$}
\affiliation{$^{a}$ Institute for Physics of Microstructures RAS, 603950, Nizhny Novgorod, GSP-105, Russia \\
$^{b}$ N.\,I.\,Lobachevsky State University of Nizhny Novgorod, Nizhny Novgorod, 603950 Russia \\
$^{c}$ Institute for Solid State Physics RAS, Acad. Ossypian 2, 142432 Chernogolovka, Moscow region, Russia\\
$^*$ E-mail: aladyshkin@ipmras.ru \\~\\}


\begin{abstract}
The spatial distribution of the differential conductance for ultrathin Pb films grown on Si(111)7$\times$7 substrate is studied by means of low--temperature scanning tunneling microscopy and spectroscopy. The formation of the quantum--confined states for conduction electrons and, correspondingly, the appearance of local maxima of the differential tunneling conductance are typical for Pb films; the energy of such states is determined mainly by the local thickness of Pb film. We demonstrate that the magnitude of the tunneling conductivity within atomically flat terraces can be spatially nonuniform and the period of the \textit{small--scale} modulation coincides with the period of Si(111)7$\times$7 reconstruction. For relatively thick Pb films we observe \textit{large--scale} inhomogeneities of the tunneling conductance, which reveal itself as a gradual shift of the quantized levels at a value of the order of 50~meV at distances of the order of 100~nm. We believe that such large--scale variations of the tunneling conductance and, respectively, local density of states in Pb films can be related to presence of internal defects of crystalline structure, for instance, local electrical potentials and stresses.
\end{abstract}



\maketitle

\section{Introduction}

Reduction in size of logical elements, sensors and conducting wires connecting them leads to that their transport properties can be influenced by discreteness of electrical charge, spatial disorder as well as quantum--size effects \cite{Ferry-book-09}.

\begin{figure}
\centering{\includegraphics[width=8cm]{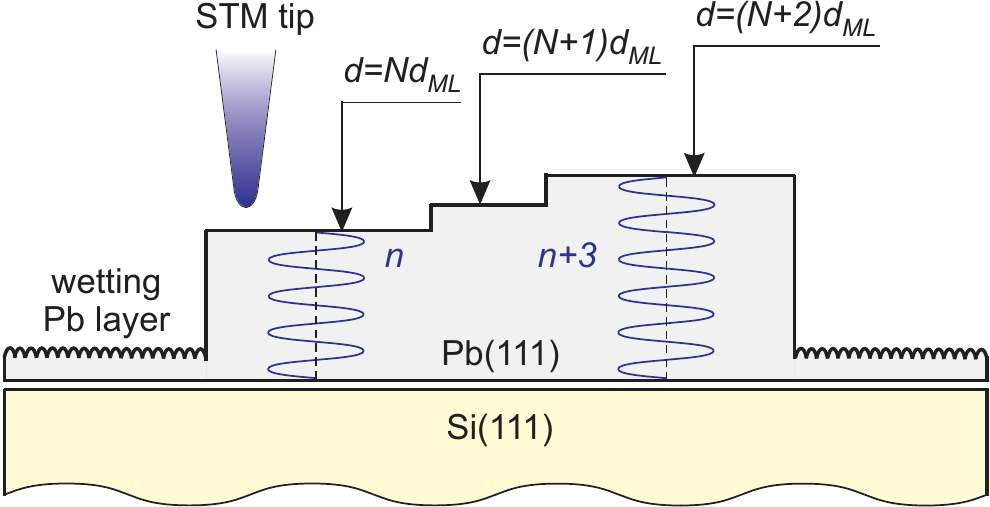}}
\caption{(color online) Schematic view of a STM tip, a disordered wetting Pb layer, a single--crystalline Pb island with atomically flat terraces as well as a spatial structure of standing electronic waves inside the island for a certain energy $E^*$ close to $E^{\,}_F$, the parameter $n$ characterises the number of the half--waves. Note that for the chosen value $E^*$ the standing waves are absent for the terrace of the local thickness $(N+1)\,d^{\,}_{ML}$.}
\label{Fig01}
\end{figure}

Ultrathin Pb films and islands appear to be convenient objects for the investigation of quantum--size effect in metallic nanostructures (\cite{Altfeder-PRL-97}--\cite{Ricci-PRB-09} and references therein). Quantized electron states can be investigated by low--temperature scanning tunneling microscopy and spectroscopy (STM/STS) \cite{Altfeder-PRL-97}--\cite{Ustavshchikov-JETPLett-2017}, transport measurements \cite{Jalochowski-PRB-88,Miyata-PRB-08} and photoemission studies \cite{Dil-PRB-06}--\cite{Ricci-PRB-09}. An appearance of peaks of the differential conductance at certain bias voltages of the sample $U_n$ and maxima of conductance at certain values of the potential difference as well as an appearance of maxima of photoemission for certain energies of photon were demonstrated for thin Pb films. In particular, the set of the energy values, corresponding of the local maxima of tunneling conductivity with respect to the Fermi level $E^{\,}_n=eU^{\,}_n+E^{\,}_F$, depend on the local thickness of Pb film and relate to the quantized energy levels in a one--dimensional potential well with boundaries at the interfaces 'metal---vacuum' and 'metal---substrate' (Fig.~\ref{Fig01}). Usually the observed effects are interpreted in terms of the resonant tunneling through quasi--stationary energy levels at $E\approx E^{\,}_n$. The energy spectrum $E^{\,}_n$ of a particle in the one--dimensional well with the constant potential is defined by the Bohr--Sommerfeld quantization rule \cite{Altfeder-PRL-02}:
    \begin{equation}
    \varphi^{\,}_1 + \varphi^{\,}_2 + 2k^{\,}_{\perp,n} d = 2\pi n.
    \label{Bohr-Sommerfeld}
    \end{equation}
Here $\varphi^{\,}_1$ and $\varphi^{\,}_2$ are the phase shifts for the electronic wave function reflected from the upper and lower interfaces, respectively; $k^{\,}_{\perp,n}$  is the spectrum of allowed values of the wave vector transverse with respect to the interfaces; $d$ is film thickness, $n=0, 1,..$ is an integer--valued index. A theory of the quantum--size effects in Pb(111) films within ab-initio models was presented in \cite{Wei-PRB-02}. It should be mentioned that besides Pb films the quantum--size effects were also observed in Ag, Cu, In, Sn and Sb \cite{Dil-PRB-06,Altfeder-PRL-04,Milun-RPP-02,Chiang-SSR-00,Komnik-67,Komnik-68}.

We would like to emphasize that the investigation of peculiarities of the resonant electron tunneling in solid--state nanostructures seems to be important diagnostic instrument similar to other techniques based on interference of waves (e.g., optical, mechanical, electronic etc.). The visualization of atomic structure of the lower interface under metallic layer was demonstrated in \cite{Altfeder-PRL-02} and this effect is based on the dependence of $\varphi^{\,}_2$ on the lateral coordinates. The possibility to visualize defects such as monoatomic steps in the substrate and the inclusions of other materials under metallic layer, which were invisible in a topographic image, was shown in \cite{Ustavshchikov-JETPLett-2017}. The estimates indicate that the thickness of a monolayer $d^{\,}_{ML}$ for the Pb(111) surface is equal to 0.285~nm, the Fermi wavelength $\lambda^{\,}_F$ is equal to 0.394~nm, therefore the ratio $\lambda^{\,}_F/d^{\,}_{ML}$ is close to 4/3 \cite{Su-PRL-01}. Consequently, the following conclusion should be valid for the electronic states in the form of standing waves with $E\simeq E^{\,}_F$ \cite{Su-PRL-01,Eom-PRL-06,Ustavshchikov-JETPLett-2017}: the energy of the state with the number of zeros $n$ for the film with the local thickness $Nd^{\,}_{ML}$ should be close to the energy of the state with the number of zeros $n+3$ for the film with the local thickness $(N+2)\,d^{\,}_{ML}$ (Fig.~\ref{Fig01}). Indeed during the measurements at the certain energy the relocation of the STM--tip from one area to another area whose local thicknesses differ in $d^{\,}_{ML}$ should result in a drastic change of the differential conductance \cite{Altfeder-PRL-98,Su-PRL-01,Ustavshchikov-JETPLett-2017}, what makes possible to reveal the areas with even or odd numbers of monolayers \cite{Ustavshchikov-JETPLett-2017}. The diagrams $U^{\,}_n-d$ allows one to recover the dependence $E(k^{\,}_\perp)$ and to estimate the thickness of the wetting layer, the effective mass and speed of electrons \cite{Altfeder-PRL-98,Su-PRL-01,Hong-PRB-09,Ustavshchikov-JETPLett-2017}. Based on the analysis of the dependence of the width of the resonant tunneling peaks on temperature, the estimates of lifetime for different scattering mechanisms were derived \cite{Hong-PRB-09}.

This paper is devoted to the experimental investigation of spatial inhomogeneity of the differential tunneling conductance for thin Pb films by means of low--temperature STM/STS. It allows us to find the correlation between local electronic properties and locations of structural defects.
It is worth to note that during STM--investigation of large areas a map of feedback--loop signal which is usually associated with a topographic image can be distorted because of different scanning speed along the $x-$ and $y-$axes and variation in temperature of a piezo--scanner, what results in uncontrolled shift of the tip along the $z-$axis. In particular, the visible height of monoatomic step in the topographic image may differ from the ideal value and the atomically flat terraces may look like not--flat. We propose a method for visualization of areas with apparent and hidden defects based on simultaneous measurements of topography and differential tunneling conductance directly during scanning process. It makes possible to discriminate the images with artifacts caused by instrumental and processing imperfections and the images with real defects. We believe that observed large--scale inhomogeneities of the differential tunneling conductance on the terraces of nominally constant height can be related with defects of crystal structure and thus point to, for instance, local electrical potentials and stresses.

\begin{figure}[t]
\centering{\includegraphics[width=8cm]{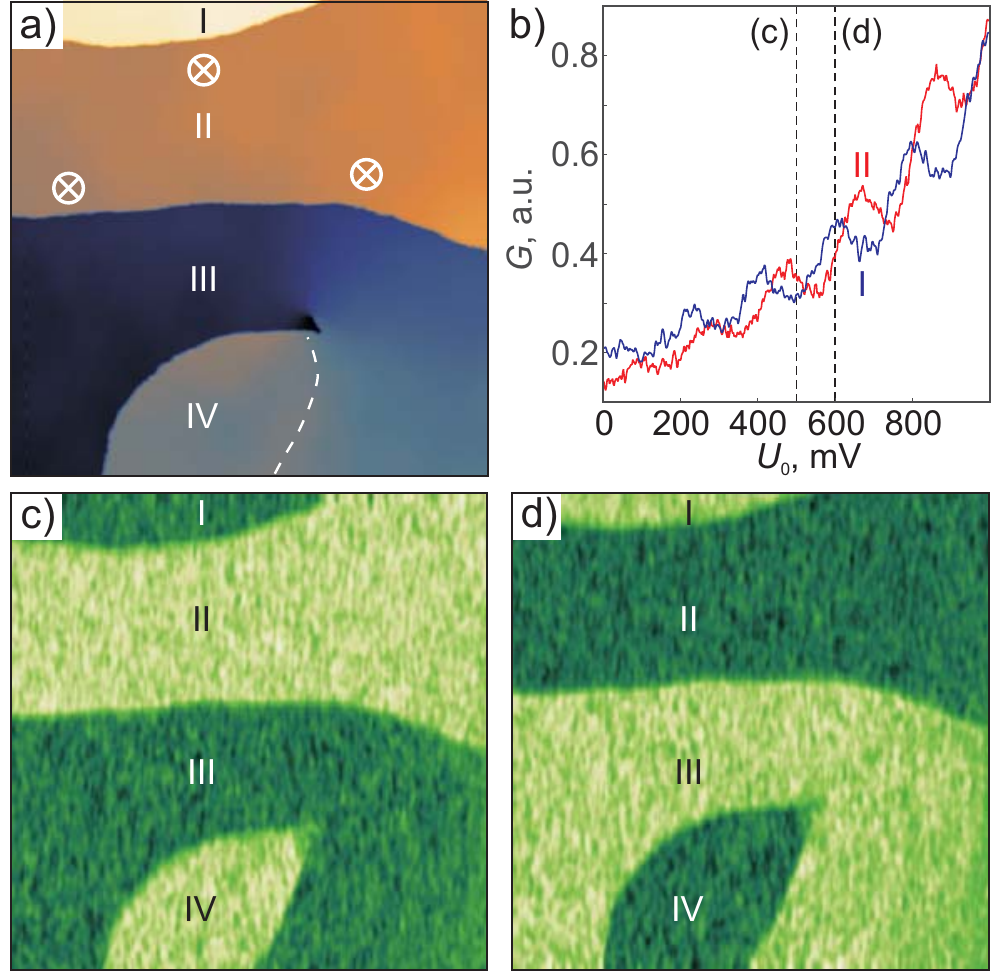}}
\caption{(color online) {\bf (a)} STM image of surface of the Pb island (175$\times$175~nm$^2$, average sample potential $U^{\,}_0=500$~mV, average tunneling current $I^{\,}_0=400$~pA), the dashed line in the lower part of the image depicts the projection of the dislocation loop on the sample surface. Hereafter the symbols $\bigotimes$ mark the reference points, which were used for the alignment of the image. {\bf (b)} The dependence $G(U^{\,}_0)$ for points within areas I and II; the vertical dashed lines correspond to the values $U^{\,}_0$ used for the maps in the panels (c) and (d).
{\bf (c,d)} The maps of the differential conductance $G(x,y,U^{\,}_0)$ for the same area shown in (a), acquired at $U^{\,}_0=500$~mV (c) and  $U^{\,}_0=600$~mV (d); $U^{\,}_1=40$~mV, $f^{\,}_0=7285$~Hz. Lighter shades correspond to higher tunneling conductance, darker shades correspond to the lower conductance}
\label{Fig02}
\end{figure}

\section{Experimental procedure}

Investigation of the electro--physical properties of Pb nanostructures is carried out on the UHV LT SPM Omicron Nanotechnology setup. Thermal deposition of Pb (Alfa Aesar, purity 99.99\%) is performed on the reconstructed surface Si(111)7$\times$7 at room temperature and pressure $3\cdot 10^{-10}$\,mbar at a rate of the order of 0.5\,nm/min, the deposition time is varied from 5 to 40\,min.

The topography of the Pb islands is studied by STM at a temperature of 78~K in the regime of constant tunneling current $I$ at a given bias potential $U$ of the sample relative to the tip of a tunneling microscope. Etched tungsten wires with apex cleaned by electron bombardment in ultra--high vacuum are used as tips. All topographic images are processed by a subtraction of a plane defined by three reference points in order to reduce global tilt. The electronic properties of the Pb islands are investigated by single point tunneling spectroscopy, consisting of measurements of series of the dependences $I(U)$ and $G(U)$ at a fixed position of the tip, where $G\equiv dI/dU$ is the differential tunneling conductivity of the tip--sample contact. In addition, maps of the local differential conductance are obtained by means of modulation scanning tunneling spectroscopy. Using a Stanford Research SR830 lock--in amplifier we measure the amplitude of the ac-component of the tunneling current, which appear for the modulated bias potential $U=U^{\,}_0+U^{\,}_1\,\cos(2\pi f^{\,}_0t)$, where $f^{\,}_0=7285$~Hz. Apparently that under the condition $U^{\,}_1\ll U^{\,}_0$ the amplitude of the oscillations of the tunneling current at the modulation frequency $f^{\,}_0$ is proportional to the differential conductance $G(U^{\,}_0)$. Provided that $f^{\,}_0$ significantly exceeds the threshold frequency of the feedback loop ($\sim200$~Hz), the modulated potential applied to the sample should not result in the appearance of any artifacts on the topographic images. Such technique \cite{Ustavshchikov-JETPLett-2017} allows us to synchronously obtain both topographic images in the regime of constant average current $I^{\,}_0$ and the dependence of $G$ on the lateral coordinates $x$ and $y$ at the given value $U^{\,}_0$.

\section{Results and discussion}

The growth of Pb on the Si(111)7$\times$7 surface at room temperature is known to occur through the Stranski--Krastanov mechanism: first a disordered wetting Pb layer with the thickness of the order of 1~nm is formed, then two-dimensional Pb islands with the upper facets corresponding to the (111) plane start to grow. It was shown \cite{Altfeder-PRL-97}--\cite{Ustavshchikov-JETPLett-2017} that the local tunneling conductance $G$ depends on $U^{\,}_0$ in non-monotonous way (Fig.~\ref{Fig02}b). Particularly, the values $U^{\,}_n$ corresponding to the peaks on $G(U^{\,}_0)$, depend on the local thickness of Pb film and on the boundary conditions for an electron wave function.

The topographic image of the surface of the Pb island (panel a) and the differential conductance maps $G(x,y,U^{\,}_0)$ at two different energies, acquired simultaneously with the topographic image at forward (panel c) and backward (panel d) scanning directions, are shown in Fig.~\ref{Fig02}. Since the interval $\Delta E$ between the neighbour maxima at the $G(U^{\,}_0)$ dependence is equal to 185~meV (Fig.~\ref{Fig02}b) and the Fermi velocity is equal to $v^{\,}_F\approx 1.8\cdot 10^8$~cm/s \cite{Altfeder-PRL-97,Ustavshchikov-JETPLett-2017}, one could estimate the local thickness of this island $d\simeq \pi\hbar v^{\,}_F/\Delta E \simeq 19$~nm or approximately 70 monolayers. The local thickness of the Pb island in the area I exceeds the local thicknesses in the areas II and III by one and two monolayers, respectively. As a consequence, the local tunneling conductances in the areas I and III are practically equal at two different energies and both differ from the conductance in the area II (Fig.~\ref{Fig02}c,d). Note that a gradual variation in the height in the vicinity of the center of the screw dislocation does not lead in a gradual change of the differential conductance. Indeed, the conductance changes drastically upon crossing a line invisible at the topographic image which corresponds to the hidden part of the dislocation loop inside the Pb film (dashed line in Fig.~\ref{Fig02}a). Since in the regions where the dislocation line is parallel to the surface such dislocation line has to be either an edge dislocation or a mismatch dislocation, the number of the monolayers changes by one upon relocation from the area III to the area IV. However in the vicinity of the dashed line the change in numbers of the monolayers occurs at constant height of the film, therefore the electronic concentration $n$ should be changed abruptly. Taking into account that $E^{\,}_F=(\hbar^2/2m^*)\,(3\pi^2n)^{2/3}$ in the model of free electron gas \cite{Ashcroft-book-79}, the bottom of the conduction band in the area IV should be shifted down at the value of the order of $\delta E^{\,}_0 \sim 2E^{\,}_F/(3N)=90$~meV in order to ensure the constancy of the Fermi level, where $m^*$ is the effective mass which is close to the mass of free electron for Pb films in the (111) direction \cite{Ustavshchikov-JETPLett-2017}, $E^{\,}_F\simeq 9.47$~eV is the Fermi energy for bulk Pb \cite{Ashcroft-book-79}, $N\simeq 70$ is the number of the monolayers in the considered area. Since $\delta E^{\,}_0$ is close to the half of $\Delta E$, the relocation across the invisible part of the dislocation loop should be accompanied by a sharp change in brightness for the maps of the tunneling conductance.

\begin{figure}[t]
\centering{\includegraphics[width=8cm]{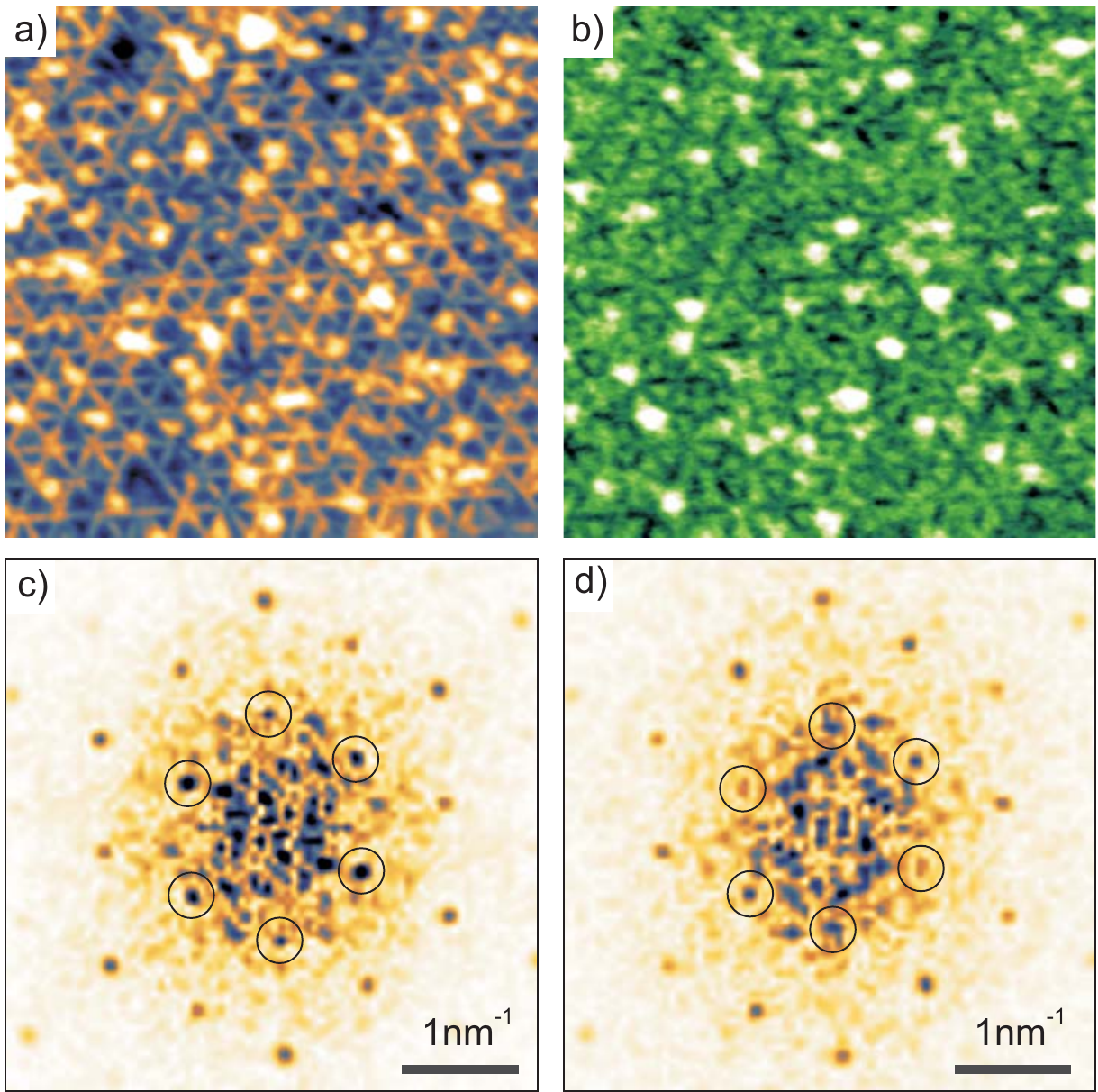}}
\caption{(color online)
(a) STM image of the surface of Pb island (35$\times$35~nm$^2$, $U^{\,}_0=500$~mV, $I^{\,}_0=200$~pA).
(b) Spatial dependence of the tunneling conductance $G(x,y,U^{\,}_0)$ at $U^{\,}_0=500$~mV, $U^{\,}_1=40$~mV.
(c,d) Amplitude of the Fourier components for the topographic image (a) and for the map of the differential conductance (b), respectively; circles depict the Fourier maxima of the first order }
\label{Fig03}
\end{figure}

\begin{figure}[t]
\centering{\includegraphics[width=8cm]{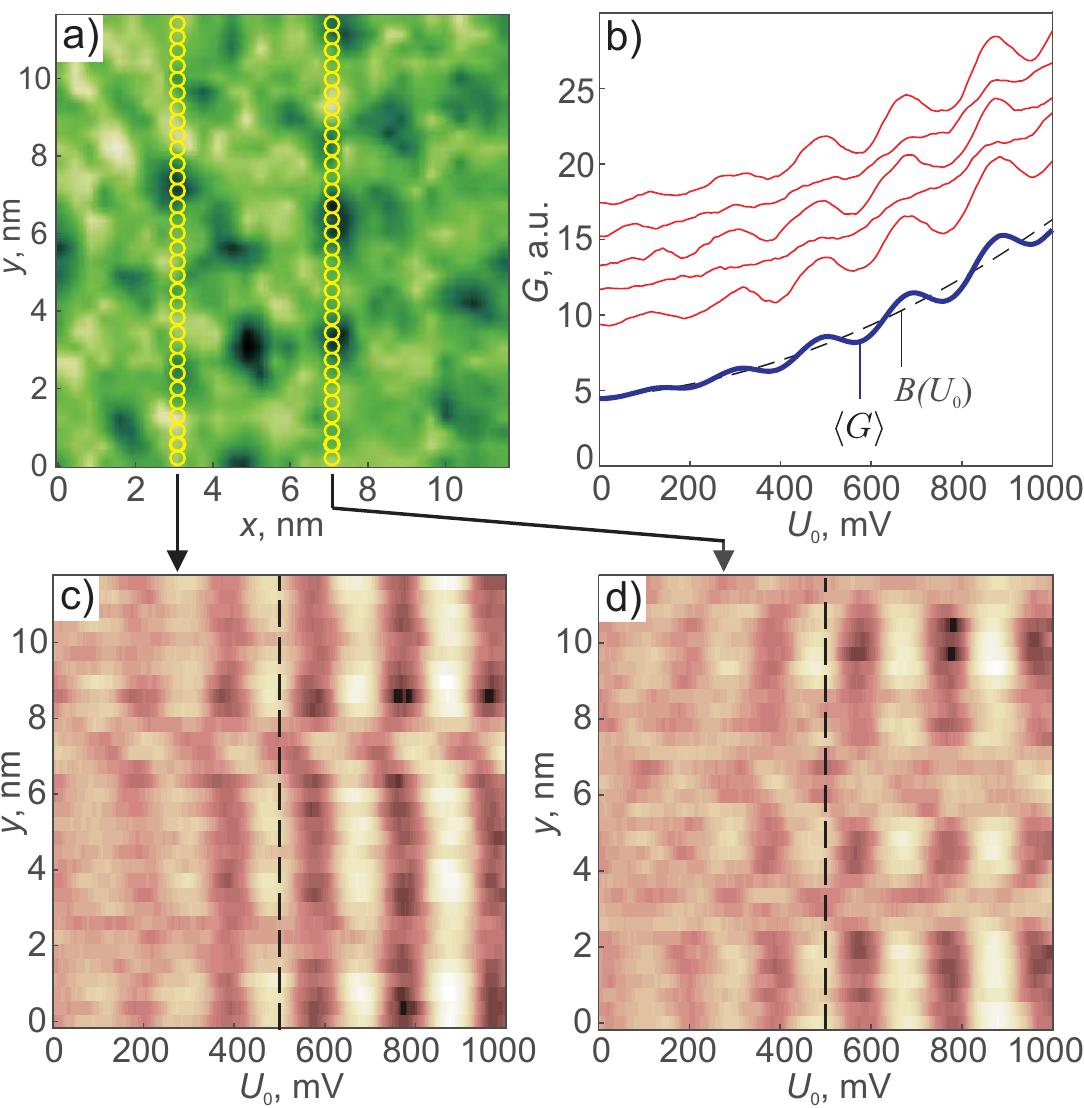}}
\caption{(color online) {\bf (a)} Map of the differential conductance $G$ for a Pb island (11.6$\times$11.6~nm$^2$, $U^{\,}_0=490$~mV, $I^{\,}_0=200$~pA). {\bf (b)} Series of the single-point spectroscopical lines $G(U^{\,}_0)$ acquired for several neighbour locations at $x=7$~nm and different $y$ values; these lines are shifted vertically for clarity. Thick blue line corresponds to the dependence $\langle G(U^{\,}_0)\rangle$; dashed line is the estimate of non-resonant background $B(U^{\,}_0)$. {\bf (c,d)} The difference of the local conductance $G(x,y,U^{\,}_0)$ and the non-resonant background $B(U^{\,}_0)$ as a function of the bias voltage $U^{\,}_0$ and the $y$ coordinate, for the locations marked in (a): $x=3$~nm (c) and $x=7$~nm (d). Brightness is proportional to the difference $G(U^{\,}_0)-B(U^{\,}_0)$. Dashed lines correspond to the $U^{\,}_0$ value, which corresponds to the map (a).}
\label{Fig04}
\end{figure}

\begin{figure}[t]
\centering{\includegraphics[width=8cm]{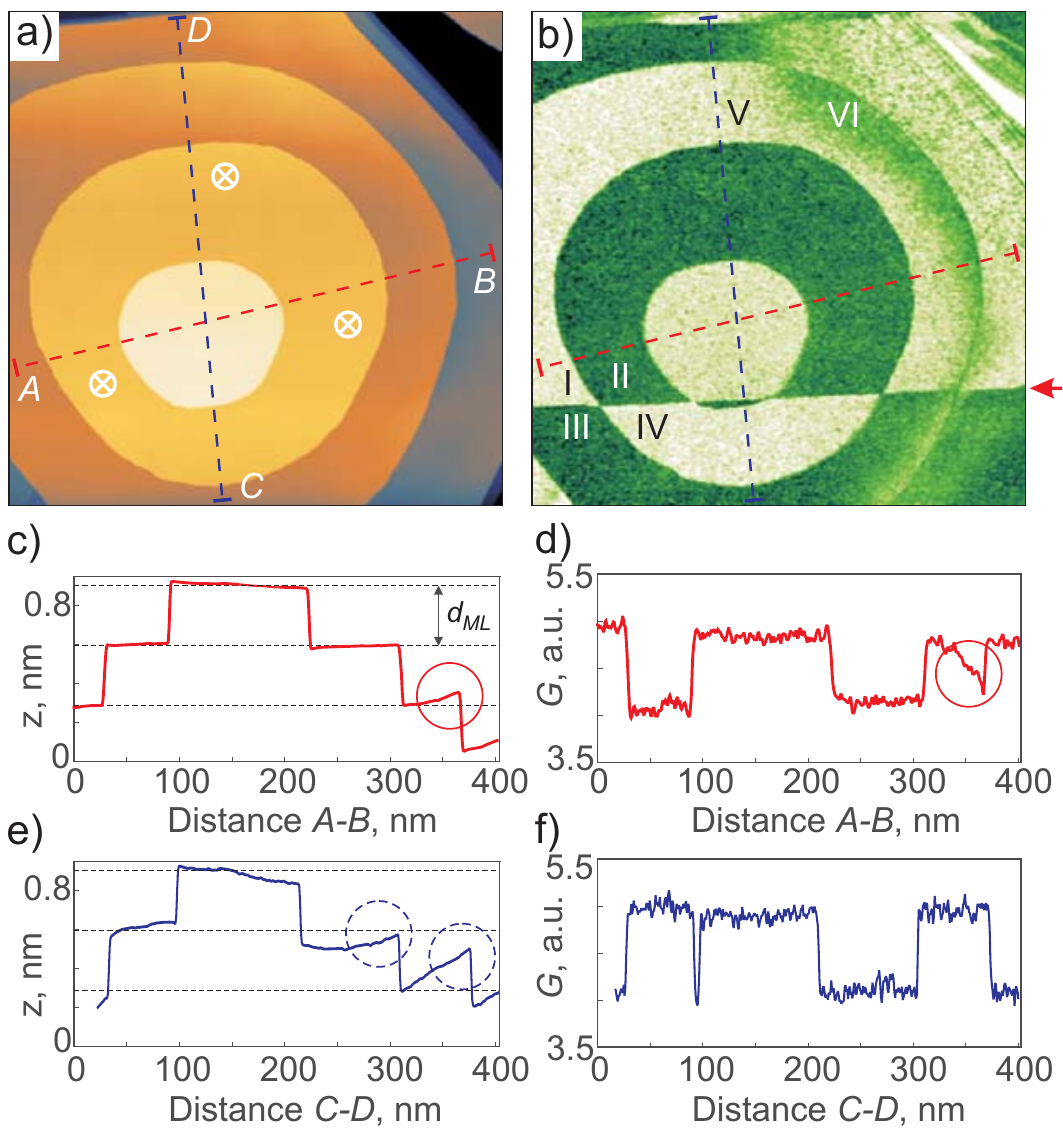}}
\caption{(color online)
{\bf (a)} STM image of the surface of Pb island (460$\times$460~nm$^2$, $U^{\,}_0=700$~mV, $I_0=400$~pA).
{\bf (b)} Map of differential conductance $G(x,y,U^{\,}_0)$ for the same area; $U^{\,}_0=700$~mV, $U^{\,}_1=40$~mV, arrow shows the position of the step of monoatomic height in the substrate.
{\bf (c,d)} Profiles of the topographic image and the differential conductance along the dash line A--B, dashed lines correspond to the heights of the Pb terraces.
{\bf (e,f)} Profiles  of the topographic image and the differential conductance along the dash line C--D, dashed circles indicate the unavoidable artifacts of processing of the topographic image}
\label{Fig05}
\end{figure}

\begin{figure*}[t]
\centering{\includegraphics[width=16cm]{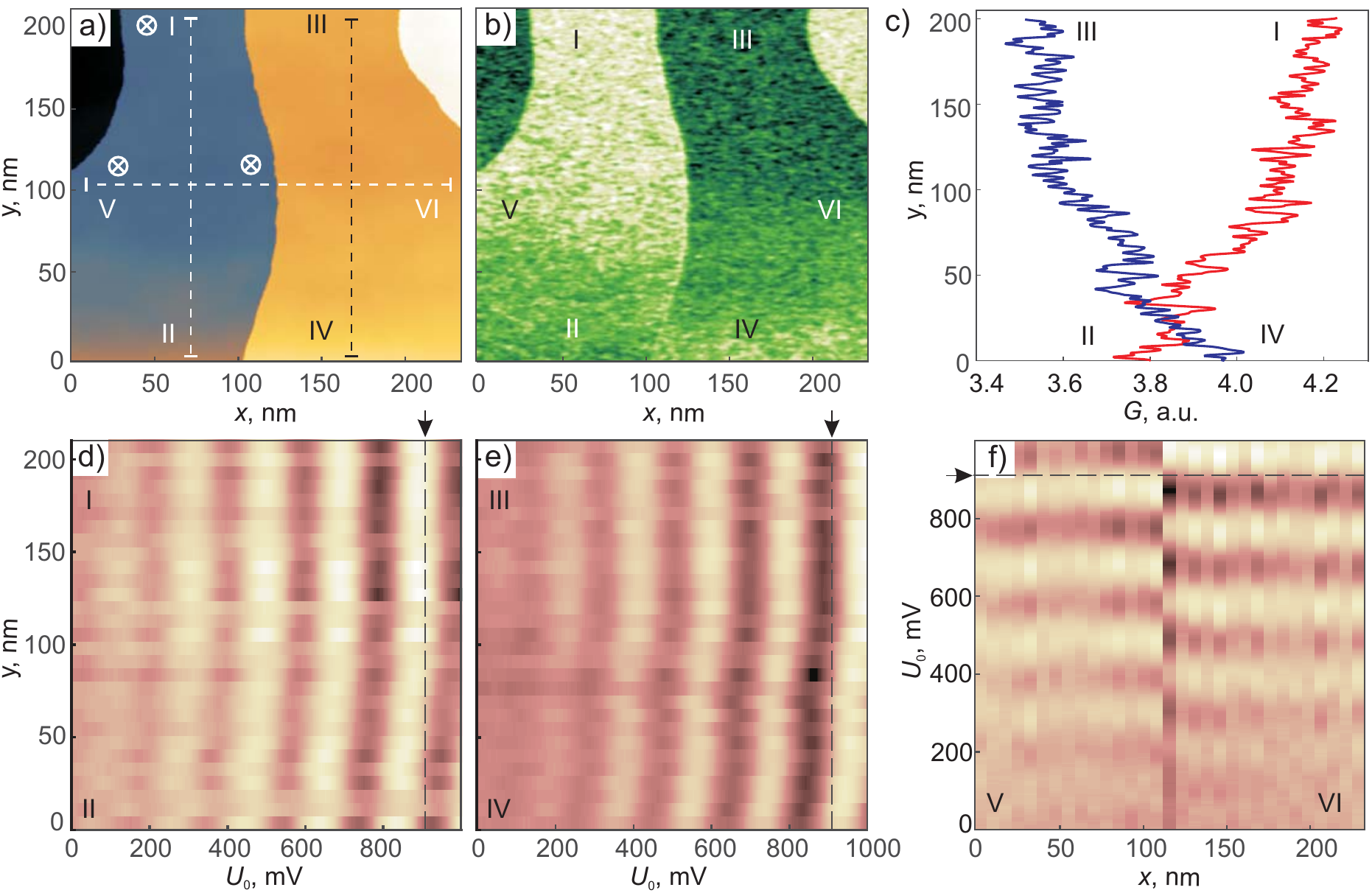}}
\caption{(color online)
{\bf (a)} STM image of the surface of Pb island(230$\times$210~nm$^2$, $U^{\,}_0=900$~mV, $I^{\,}_0=200$~pA). The three points marked in the figure used for levelling.
{\bf (b)} Map of the differential conductance $G(x,y)$ for the same area acquired at $U^{\,}_0=900$~mV, $U^{\,}_1=40$~mV.
{\bf (c)} Profiles of the differential conductance along the vertical lines I--II and III--IV.
{\bf (d,e)} Difference of the local conductance $G(x,y,U^{\,}_0)$ and the non-resonant background $B(U^{\,}_0)$ as a function of the bias $U_0$ and the $y-$coordinate along the lines I--II (d) and III--IV (e). Brightness is proportional to the value $G(U^{\,}_0)-B(U^{\,}_0)$. Vertical dash lines correspond to the value $U^{\,}_0$ used for the acquisition of the map in (a).
{\bf (f)} Difference of the local conductance $G(x,y,U^{\,}_0)$ and the non-resonant background $B(U^{\,}_0)$ as a function of bias $U_0$ and the $x-$coordinate along the line V--VI.}
\label{Fig06}
\end{figure*}

Note that the tunneling conductance even within single terrace is not strictly constant. Figure~\ref{Fig03} shows the topography (a) and the map of conductance (b) for the Pb island and it has the thickness about six monolayers over the wetting layer according to our estimates. This topographic image contains signatures of hexagon lattice what is confirmed by presence of well-defined peaks of the first and second orders on the map of amplitudes of the Fourier components (Fig.~\ref{Fig03}c). The spatial modulation with the same wave vectors takes place for the tunneling conductance (Fig.~\ref{Fig03}b and d). One can suggest that the observed periodicity is conditioned by an influence of the crystalline structure on the tunneling density of states in Pb films, since the period of modulation coincides with the period of the reconstruction Si(111)7$\times$7. Such effect can be related with the variation of the phase of electron wave reflected from the interface 'metal--substrate' at different points of the surface and can reveal itself as a periodic shift of the peaks of the tunneling conductance as well as the Moir\'{e} contrast \cite{Altfeder-PRL-98,Altfeder-PRL-02}.

For a more careful investigation of the short--scale inhomogeneities of the differential conductance we consider the area 11.6$\times$11.6~nm$^2$ of atomically-flat terrace of the Pb island with a thickness of 60--70 monolayers. For this island we perform the series of measurement at the grid 32$\times$32 ({\it grid spectroscopy}) with a step of 0.36~nm. The spatial dependence of the conductance on the coordinates $x$ and $y$ at $U^{\,}_0=490$~mV is shown in Fig.~\ref{Fig04}a. Several typical local dependences $G(U_0)$ are shown in Fig.~\ref{Fig04}b. Note that depending on the measurement location the local tunneling spectra contain either the set of well-defined peaks or these peaks are badly distinguishable. For analysis of the dependence of the position and amplitude of the resonant peaks on energy and coordinates it is convenient to remove a non-resonant background. With this purpose all 1024 spectral curves were averaged over the scanning area and then the mean conductance $\langle G(U^{\,}_0)\rangle$ (the thick solid line in Fig.~\ref{Fig04}b) was approximated by a third-order polynomial dependence in order to exclude any effects of quantum--confined states. This approximation polynomial $B(U^{\,}_0)$ ({\it background}) is shown in Fig.~\ref{Fig04}b by dashed line. The differences of the local conductance from the non-resonant background $B(U^{\,}_0)$ as a function of bias $U^{\,}_0$ and the coordinate $y$ for two values $x=3$~nm (c) and $x=7$~nm (d) are shown in Fig.~\ref{Fig04}. It is easy to see that the areas with pronounced peaks of the differential conductance alternate with the areas with no peaks, unclear peaks or peaks shifted to different energy.

Figure~\ref{Fig05} shows the topographic image and the map of the differential conductance for the Pb islands, where the terraces of monoatomic height have a form of concentric circles. It should be noted that the conductance in the areas I and IV is close to the maximal value, while the conductance in the areas II and III is close to the minimal value. It allows us to conclude that there is an invisible step of monoatomic height in the substrate what results in a drastic change in conductance within single Pb terrace (paths I--III and II--IV in Fig.~\ref{Fig05}b). Besides that we found out the terraces with a gradual change in conductance at a given energy (for instance, paths V--VI in Fig.~\ref{Fig05}b and I--III and II--IV in Fig.~\ref{Fig06}b). The appearance of the regions with gradual change of differential conductance looks quite surprising since in elementary models a film thickness should be equal to an integer number of monolayers and, consequently, the tunneling conductance should varies discretely. Note that the appearance of the gradual contrast at the maps $G(x,y,U^{\,}_0)$ cannot be related to a modification of a tip apex during measurements, since the areas with sharp and edges are observed simultaneously. Figure~\ref{Fig05}c, d show the cross-sections of the topographic image and the map of the differential conductance along the A--B line, which is close to the direction of fast scanning. It is easy to see that the gradual change of conductivity marked by the circle in Fig.~\ref{Fig05}d corresponds to the gradual variation of the height of the order $0.2\,d^{\,}_{ML}$ at the topographic image marked by the circle in Fig.~\ref{Fig05}c.

We think that the observed phenomenon is related to the presence of internal stress in Pb film which affect both the actual height of the terraces and the energy of the bottom of the conduction band. To the contrary, the cross-section of the map of the differential conductance along the C--D line close to the direction of slow scanning is a function which has two limiting values (Fig.~\ref{Fig05}f). As a consequence, the local thickness of the Pb film along this line should be varied in a quantized way and the complicated shape of the profile along the same line (dashed circles in Fig.~\ref{Fig05}e) is apparently an artifact caused by imperfection of both piezo--scanner and procedure of the compensation of the global tilt.

In order to study the peculiarities of the differential conductance for Pb films with gradual large--scale inhomogeneities as a function of the coordinates and the energy we investigate the area of the Pb island with three monoatomic steps; the height of this island is about 60 monolayers. The topography of this islands is shown in  Fig.~\ref{Fig06}a. A detailed analysis of the cross-sections along the lines I--II and III--IV profiles points to a monotonous variation of the height of the terraces at about $0.2\,d^{\,}_{ML}$ in the interval from $y=0$ to $y=100$~nm, what can be easily recognized in Fig.~\ref{Fig06}a by a change of colors. The map of the differential conductance (Fig.~\ref{Fig06}b) evidences about the presence of sharp boundaries, for instance, upon relocating from the area I to the area III, whose heights differ by one monolayer. However upon the tip relocation from the area I to the area II (or III--IV) a gradual change of the tunneling conductance takes place: the conductance at $U^{\,}_0=900$~mV decreases along the line I--II and it increases along the line III--IV (Fig.~\ref{Fig06}b,c). In the same area the series of the current -- voltage characteristics and the spectra of the differential tunneling conductance are obtained at the grid 32$\times$32 and then a non-resonant background is subtracted using the procedure described above. The results of these measurements indicate that there is the gradual shift of the levels of the quantum-confined states towards higher energy of the order of 50~mV happens along the $y-$axis (Fig.~\ref{Fig06}d,e). In the other words, we observe the gradual transition from the local maximum on the $G(U^{\,}_0)$ dependence at the energy 900~mV (vertical line in Fig.~\ref{Fig06}d) to the local minimum provided that the tip is shifted along the line I--II, what corresponds to the decrease in the tunneling conductance (Fig.~\ref{Fig06}b,c). Similarly the shift along the line III--IV at the energy 900~mV causes the gradual increase of the conductance (Fig.~\ref{Fig06}e). It would be noted that upon relocation in the horizontal direction between the areas V and VI a position-independent differential conductivity is observed with abrupt change at the terrace edge (Fig.~\ref{Fig06}f). Consequently, the monotonous change of the height of the terrace is accompanied by the changes of electron properties of the sample and results in a systematic shift of the quantized quantum--confined levels in the interval from $y=0$ to $y=100$~nm. Note that the observed shift of the energy levels is close to the estimate of the shift of the bottom of the conduction band $\delta E^{\,}_0$ caused by the change of the local electron density.

Turning back to the simplest model (\ref{Bohr-Sommerfeld}) of the localised electron states in a one-dimensional potential well, one can state that the gradual shift of the quantum--confined levels can be caused by, first, a monotonous change in the thickness of Pb layer $d(x,y)$, second, a change in the energy of the bottom of the conduction band $E^{\,}_0(x,y)$ and, third, a change in the boundary conditions at the interface 'metal---substrate'. The last circumstance is probably responsible for the small--scale inhomogeneity of the electronic properties. We suppose that mechanical stress of the crystalline structure, what arise during growth process of the Pb structures and can result in the change both the energy $E^{\,}_0$ and the height of terraces, is the most probable origin of appearing of the areas with the gradual inhomogeneity of the tunneling conductance.

\section{Conclusion}

We demonstrate that the change in the local thickness of the Pb film by one monolayer due to monoatomic steps at the lower or upper interfaces results in abrupt spatial variation (with typical length scale of the order of several nm) of the average differential conductance at the given energy. The observed small--scale modulation of the tunneling conductance (with typical period $\sim3$~nm) is related with an influence of the periodic potential of the substrate, which is the reconstruction Si(111)7$\times$7. Besides that the large-scale variations of the differential tunneling conductance within single terrace of the Pb island are observed, manifesting as the gradual change of the quantum-confined energy levels at a value of the order of 50~nm at the lateral scales of the order of 100~nm. A possible reason of the appearance of the large--scale variations is the spatially--inhomogeneous internal stress in thin Pb films, which can result in {\it non-quantized} changes in the thickness of the Pb layer different from the integer number of monolayers. Systematic investigation of the dependence of differential conductivity on the coordinates and energy is a convenient method for studying of internal defects in Pb nanostructures.

\section{Acknowlednements}

The authors thanks D.\,Yu. Roditchev and A.\,N. Chaika for fruitful discussions. The work was performed with the use of equipment at the Common Research Center 'Physics and Technology of Micro- and Nanostructures' at Institute for Physics of Microstructures RAS (IPM RAS). The work was supported partly by the Presidium of RAS under the program 0035-2018-0019 (sample preparation), partly by the RFBR grant 19-02-00528 (STM-STS measurements), partly by the Governmental program for IPM RAS in 2019 (interpretation of results) and partly by the Governmental program for the Institute for Solid State Physics RAS in 2019 (interpretation of results).

\end{document}